\documentclass[aps,prb,twocolumn,showpacs,amsmath,amssymb,superscriptaddress,floatfix,showkeys]{revtex4}	
\usepackage{epsfig}
\usepackage[latin9]{inputenc}
\usepackage{color}
\usepackage{graphicx}
\usepackage{hypernat}
\usepackage{hyperref}

\def\etal{{\em{et al.}}}
\newcommand{\longoverbrace}[2]{\overbrace{#1}^{\text{\hbox to 0cm{\hss #2 \hss}}}}  
\newcommand{\longunderbrace}[2]{\underbrace{#1}_{\text{\hbox to 0cm{\hss #2 \hss}}}}  

\begin{document}


\title{Effective Cluster Typical Medium Theory for Diagonal Anderson Disorder Model in One- and Two-Dimensions}

\author{Chinedu E. Ekuma}
\altaffiliation{Electronic Address: cekuma1@lsu.edu}
\affiliation{Department of Physics \& Astronomy Louisiana State University,
Baton Rouge, LA 70803, USA}
\affiliation{Center for Computation and Technology, Louisiana State University, Baton Rouge, LA 70803, USA}

\author{Hanna Terletska}
\affiliation{Department of Physics \& Astronomy Louisiana State University,
Baton Rouge, LA 70803, USA}
\affiliation{Brookhaven National Laboratory, Upton, New York 11973, USA}

\author{Zi Yang Meng}
\affiliation{Department of Physics \& Astronomy Louisiana State University,
Baton Rouge, LA 70803, USA}
\affiliation{Center for Computation and Technology, Louisiana State University, Baton Rouge, LA 70803, USA}

\author{Juana Moreno}
\affiliation{Department of Physics \& Astronomy Louisiana State University,
Baton Rouge, LA 70803, USA}
\affiliation{Center for Computation and Technology, Louisiana State University, Baton Rouge, LA 70803, USA}

\author{Mark Jarrell}
\altaffiliation{Electronic Address: jarrellphysics@gmail.com}
\affiliation{Department of Physics \& Astronomy Louisiana State University,
Baton Rouge, LA 70803, USA}
\affiliation{Center for Computation and Technology, Louisiana State University, Baton Rouge, LA 70803, USA}

\author{Samiyeh Mahmoudian}
\affiliation{Department of Physics, Florida State University, Tallahassee, FL 32301, USA}

\author{Vladimir Dobrosavljevi\'{c}}
\affiliation{Department of Physics, Florida State University, Tallahassee, FL 32301, USA}

\date{\today}

\begin{abstract}
\noindent We develop a cluster typical medium theory to study localization in disordered electronic
systems. Our formalism is able to incorporate non-local correlations beyond the local typical 
medium theory in a systematic way. The cluster typical medium theory utilizes the momentum
resolved typical density of states and hybridization function to characterize the localization 
transition.  We apply the formalism to the Anderson model of localization in one- and two-dimensions. 
In one dimension, we find that the critical disorder strength scales inversely with the linear cluster size 
with a power-law, $W_c \sim (1/L_c)^{1/\nu}$; whereas in two dimensions, the critical disorder strength 
decreases logarithmically with the linear cluster size. Our results are consistent with previous 
numerical work and in agreement with the one-parameter scaling theory. 
\end{abstract}



\maketitle

\section{Introduction}
\label{sec:intro}
Over the past several decades, disorder-driven (Anderson) localization has been a subject of intensive 
theoretical\cite{PhysRevLett.42.673,PhysRevLett.47.1546,PhysRev.109.1492,Lagendijk2009,
RevModPhys.57.287,PhysRevLett.82.382,PhysRevB.81.155106,Pichard1981,
Vlad2003,RevModPhys.46.465,PhysRevB.76.045105,Bhatt2012,Gorkov1979} and experimental \cite{PhysRev.154.750,Morgan1971,
Kramer1993,Sharvin1982,Bergmann1984,PhysRevB.50.8039,Kravchenko2004,RevModPhys.73.251} studies. 
It has recently been realized experimentally with atomic matter waves.\cite{Billy2008,Roati2008,PhysRevLett.101.255702} 
In the original seminal paper of Anderson, \cite{PhysRev.109.1492} localization is defined as the absence 
of diffusion.  It generally arises from quantum interference between different particle trajectories and 
depends strongly on the dimensionality of the system. According to the one-parameter scaling theory, 
\cite{PhysRevLett.42.673} there is no delocalized phase in one- and two- dimensions, whereas, 
a metal-insulator-transition (localization/delocalization phase) occurs at finite disorder strength in three-dimensions 
(3D).
Effective medium theories like the coherent potential approximation 
(CPA) \cite{RevModPhys.46.465} and its cluster extensions, including the dynamical cluster approximation 
(DCA), \cite{PhysRevB.63.125102,RevModPhys.77.1027,PhysRevB.61.12739}
average over disorder, favoring a metallic solution, and fail to
capture the localized state in 1 and 2D, and even the transition in 3D.

Here, we develop an extension of the Typical Medium Theory~\cite{Vlad2003} which is able to capture the 
localized state. The study of disordered systems rely on probability distribution functions 
(PDFs) to measure `random' quantities of interest.  To characterize localization, the most important 
quantity in the majority of physical or statistical problems is usually 
the ``typical'' value of the `random' variable, which corresponds to
the most probable value of the PDF. \cite{mvp1} 
In most systems the nature of the PDF is not known a priori; as such, we have limited 
information via the moments or cumulants of the PDFs. Under such situation, the ``typical'' or the most probable 
value of the PDF \cite{mvp1} contains 
important and direct information.  Different from some systems where the first moment (the arithmetic 
average) is a good estimate of the random variable, the Anderson localization is a non-self-averaging 
phenomenon. Close to the critical point the electronic quantities fluctuate strongly  and 
the corresponding PDF of the local density of states is very asymmetric with long 
tails \cite{PhysRevLett.72.526,Janssen1994} such that infinitely many moments are needed to describe 
it.\cite{mvp1} In some cases, the corresponding moments might not even exist 
especially close to the critical point. \cite{PhysRevLett.94.056404} 

The arithmetic average of random one-particle quantities is not critical at the Anderson localization 
transition.  This is the reason that most mean field theories like the CPA \cite{RevModPhys.46.465} and its cluster extensions
including the DCA, \cite{PhysRevB.63.125102,RevModPhys.77.1027,PhysRevB.61.12739} fail to provide a proper description 
of Anderson localization in disordered systems. This failure is intrinsic to these theories as the 
algebraically averaged quantities, i.e., averaged density of states, used in these methods always favor 
the metallic state.  This can be understood from the fact that in an infinite system with localized states, 
the average density of states will remain continuous while the local density of states become discrete. 
See for e.g., Refs. \onlinecite{Thouless1974,Thouless1970,PhysRevLett.72.526,Lagendijk2009,
Janssen1998,PhysRevB.74.153103,Vlad2003} for a detailed discussion.

In contrast to the arithmetic average, the geometrical average, 
\cite{Vlad2003,Janssen1998,Janssen1994,PhysRevLett.72.526,Crow1988} gives a better approximation 
to the most probable value of the local density of states.  \citet{Vlad2003} proposed the typical 
medium theory (TMT) to study disorder system, where the arithmetically averaged local density of 
states is replaced with the typical density of states (TDOS), where the geometrical average is used. 
They demonstrated that the TDOS vanishes continuously as the disorder strength increases towards 
the critical point, and it can be used as an effective mean field order parameter for the Anderson 
localization.  

However, the TMT proposed in Ref.~\onlinecite{Vlad2003} is a single-site self-consistent mean field 
theory. Due to its local nature, it fails to capture the effect of spatial correlations.   
In this work, we extend the local typical medium theory \cite{Vlad2003} to a cluster version 
utilizing the ideas from the  dynamical cluster approximation.\cite{PhysRevB.63.125102,RevModPhys.77.1027,
PhysRevB.61.12739} The DCA systematically incorporates spatial inter-site correlations. 
Therefore quantum coherence, which is important for Anderson localization, is captured. 
This is the key motivation of our work.  A concomitant motivation is that recent theoretical 
work shows that rare events \cite{Bhatt2012} play an important role in Anderson localization; 
our formalism might be able to take rare events into consideration.

We develop a cluster typical medium theory (CTMT) for studying disorder systems. 
Our cluster self-consistent mean field theory systematically incorporates non-local 
correlation effects into the local TMT (for a review of cluster approximation see 
Refs.~\onlinecite{agonis,RevModPhys.77.1027}). In our formalism, the typical density of states (TDOS) 
is calculated using the geometrical average of the imaginary part of the Green function. 
We utilize the TDOS and the hybridization rate 
to characterize the localization transition. 
The imaginary part of the hybridization rate measures the hopping amplitude or diffusion of 
electrons from the cluster to the typical medium  (c.f.\ Fig.~\ref{algorithm-self_consistency} (a)). 
Then, the point where this quantity vanishes corresponds to the absence 
of diffusion and the onset of Anderson localization.   We also note that the TDOS 
vanishes at the same point as the hybridization rate.  We apply the developed 
CTMT approach to study the non-interacting Anderson model in one and two dimensions, and 
we briefly describe the failure of the method in three dimensions. For a review of the progress 
in lower dimensional Anderson localization, see for e.g., Refs.~\onlinecite{Ishii1973,RevModPhys.57.287,
PhysRevLett.45.842,Thouless2010,Kramer1993,MacKinnon1983,Wischmann}. 
Our results show that the CTMT provides a proper mean field description of Anderson 
localization for lower dimensional systems. While the present study is for diagonal Anderson disorder model,
the method can easily be extended to other disorder
distributions like Gaussian, binary disorder, etc.

The rest of this paper is organized as follows. After Section ~\ref{sec:intro}, 
the basic formalism and description of the CTMT self-consistency is provided in Section~\ref{sec:formalism-theory}. 
Section~\ref{sec:results} shows our computed results. We conclude  in Section~\ref{sec:summary}.


\section{Model and Formalism}
\label{sec:formalism-theory}
We consider the Anderson model \cite{PhysRev.109.1492} 
with a diagonal on-site random disorder potential. The Hamiltonian is given by 
\begin{equation} \label{eqn:1}
\hat{H}
=
- t\sum_{<i,j>} (c_{i}^\dagger c_{j} + c_{j}^\dagger c_{i}) + 
\sum_{i} (V_i - \mu) n_{i}.
\end{equation}
The operators $c_{i}^\dagger$($c_{i}$) create (annihilate) a quasi-particle in a Wannier orbital on site $i$ and 
$n_{i} = c_{i}^\dagger c_{i}$ is the number operator, $\mu$ is the 
chemical potential, and $t$ is the hopping matrix element between nearest-neighbor $<i,j>$,  
which we set $4t = 1$ as the energy unit. The local 
potentials $V_i$ $\in$ $[-W,+W]$ are randomly distributed according to a probability distribution $P(V_i)$ with a 
box distribution function: 
\begin{equation} \label{eqn:2}
P(V_i)
=
\frac{1}{2W} \Theta (W - | V_i |)
\end{equation}
where the strength of the disorder in units of $4t$ is parametrized by the width $W$ of the box, and $\Theta(x)$ 
is the step function.  
\begin{figure}[b]
\begin{center}
 \includegraphics[trim = 0mm 0mm 0mm 0mm,width=1\columnwidth,clip=true]{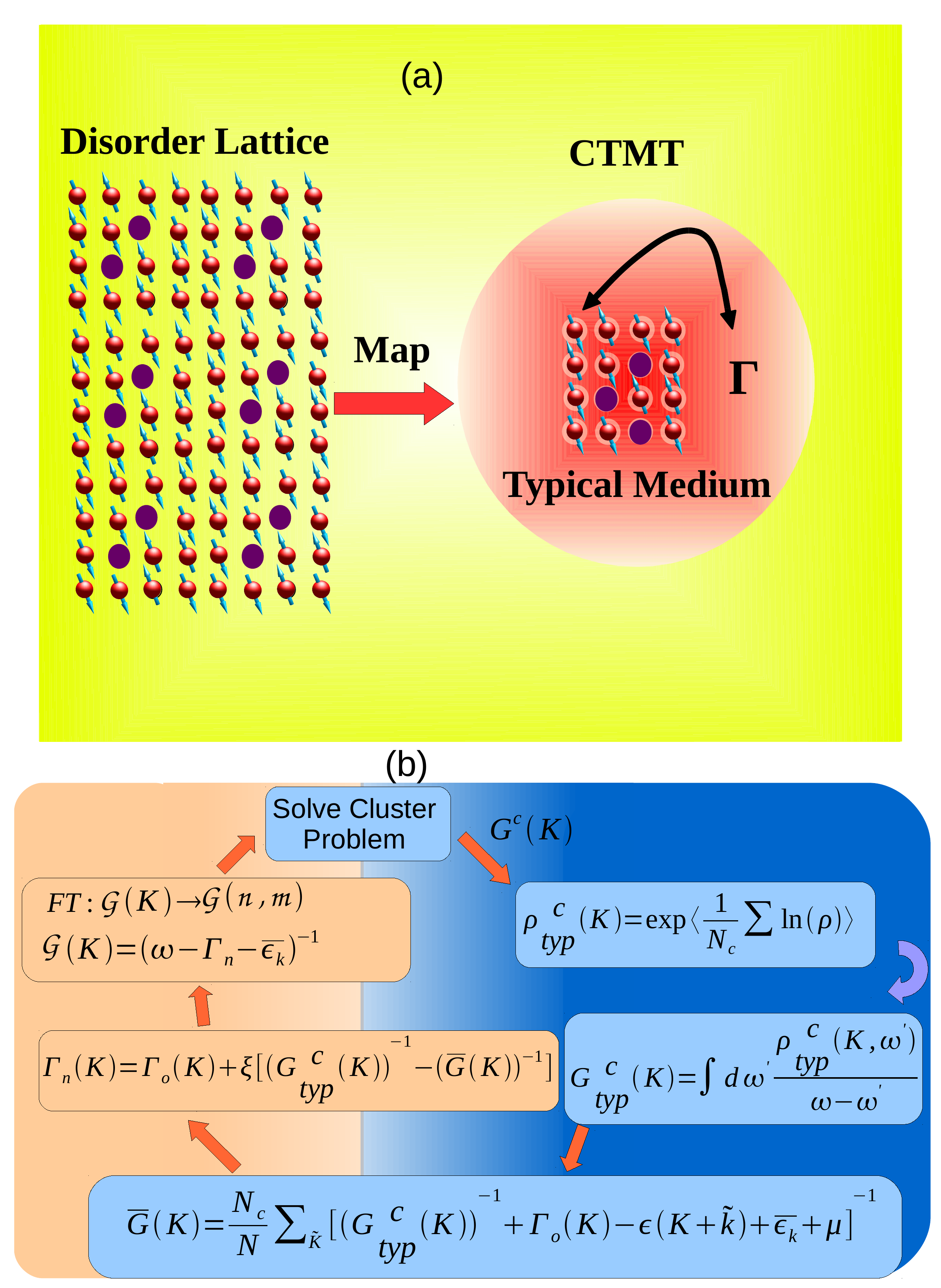}
\caption{(Color online) (a) A schematic one-dimensional diagram of the 
environment of the cluster typical medium theory (CTMT). This diagram depicts the mapping of 
a disordered, infinite lattice to a finite cluster self-consistently embedded in the typical 
effective medium. (b) The self-consistent loop of the computational procedure of the cluster typical 
medium theory.}
\label{algorithm-self_consistency}
\end{center}
\end{figure}
 
The cluster typical medium theory (CTMT) combines the self-consistent frameworks
of the DCA \cite{PhysRevB.63.125102} and the single site TMT approaches. \cite{Vlad2003} In particular, the 
CTMT maps the given disordered lattice system into a finite cluster which is embedded in an effective 
self-consistent typical medium (c.f.\ Fig.~\ref{algorithm-self_consistency}(a)). Note that unlike in the 
usual DCA scheme where the effective medium is constructed via algebraic averaging over disorder 
configurations, in the CTMT scheme, geometric averaging is used. By mapping a $d$-dimensional lattice 
containing $N$ sites to a finite small cluster containing $N_c$ $=$ $L^d_c$ sites, where $L_c$ is the linear 
dimension of the cluster, we dramatically reduce the computation effort.\cite{RevModPhys.77.1027} 
Unlike the single-site methods commonly used to study disordered systems, such as the coherent potential 
approximation (CPA) \cite{RevModPhys.46.465,PhysRev.156.809} or the local TMT, \cite{Vlad2003} the CTMT 
ensures that non-local, spatial fluctuations, which are neglected in single-site approaches,  are systematically 
incorporated as the cluster size $N_c$ increases. Short length scale correlations inside the cluster are
treated with exact numerical methods such as Monte Carlo (MD) or exact diagonalization, while  long length 
scale correlations are treated within the typical medium (c.f.\ Fig.~\ref{algorithm-self_consistency}(a)). 
At the limit of cluster size N$_c$ = 1, the CTMT recovers the local TMT, and at the limit of 
N$_c$ $\rightarrow$ $\infty$, the CTMT becomes exact. Hence, between these two limits, the CTMT 
systematically incorporates non-local correlations into the local TMT.

Let us describe our formalism in detail. In the DCA, the average 
density of states (DOS) at each cluster momentum $K$ is defined as
\begin{equation} \label{eqn:3}
\rho_{avg}^c(K,\omega)= \langle \rho^c(K,\omega)\rangle = -\frac{1}{\pi} \langle \textnormal{Im}  G^c (K,K,\omega) \rangle,
\end{equation}
where the superscript `$c$' denotes cluster, and $\left\langle \right\rangle$ is the disorder average.
For a single site $N_c = 1$, we recover the CPA. The $K$-dependent cluster 
Green function is obtained from the site dependent Green function $G_c(i,j,\omega)$ via 
\begin{equation}
G^c (K,K,\omega) = \frac{1}{N_c} \sum_{i,j} e^{iK\cdot (R_i-R_j)} G_c(i,j,\omega).
\end{equation}

In the CTMT, we first obtain 
$\rho^{c}(K,\omega)=-\textnormal{Im} G^c (K,K,\omega)/\pi$ for each cluster disorder 
configuration.  It is easy to show, using the Lehmann representation,\cite{Gross1991,fetter1971quantum} 
that $\rho^{c}(K,\omega)\geq 0$ for each $K$, $\omega$, and disorder configuration.  
We then calculate using geometric averaging the cluster-momentum-resolved typical density of states (TDOS) for 
each $K$ as 
\begin{equation} 
\rho_{typ}^c(K,\omega)
=
\exp \left\langle \ln \rho^c (K,\omega) \right\rangle.
\label{eq:geometric_rho}
\end{equation}
To ensure the causality of the Green function, we carry out the Hilbert transformation of the TDOS 
to obtain the typical cluster Green function as 
\begin{equation} 
G_{typ}^c(K,\omega)
=
\int d \omega' \frac{\rho_{typ}^c(K,\omega')}{\omega - \omega'}.
\label{eq:Hilbert}
\end{equation}
And we use the typical Green function to continue the self-consistency loop. 
A schematic diagram of the CTMT self-consistency loop is shown in Fig.~\ref{algorithm-self_consistency}(b). 
We have adopted the hybridization rate function $\Gamma (K,\omega)$ as the order
parameter to control the convergence. This stems from the fact that $\textnormal{Im}$($\Gamma(K,\omega))$
measures the rate of electron hybridization between the cluster and the typical medium. When 
$\textnormal{Im}$($\Gamma(K,\omega)$) is zero, the hopping between cluster and medium vanishes and the 
electrons in the system are localized. We  note that at this point the TDOS also becomes zero, 
as such, the TDOS and $\textnormal{Im}$($\Gamma(K,\omega)$) act as mean field order parameters within 
our effective medium for detecting Anderson localization. The CTMT iterative procedure is described as 
follows: 
\begin{enumerate}
 \item 
We start by proposing an initial hybridization function $\Gamma_o (K,\omega)$, 
where the subscript $o$ denotes old.  The choice of the starting guess for the hybridization function may 
be based on a priori knowledge, i.e., in case we have information about the self-energy $\Sigma(K,\omega)$ 
and cluster Green function $G^c(K,\omega)$, $\Gamma_o (K,\omega)$ can be calculated as
\begin{multline} \label{eqn:oldhybrid}
\Gamma_o (K,\omega) = \omega - \overline{\epsilon}(K) - \Sigma (K,\omega) - 1/G^c(K,\omega)
\end{multline}
where $\overline{\epsilon}(K)$ $=$ $\displaystyle \frac{N_c}{N}\sum_{\tilde{k}}{\epsilon(K+\tilde{k})}$ is the 
coarse-grained bare dispersion with $\tilde{k}$ summed over $N/N_c$ momenta 
inside the cell centered at the cluster momentum $K$.\cite{PhysRevB.61.12739}   However, if nothing is 
known a priori, the guess $\Gamma_o (K,\omega) \equiv 0$ may serve as the starting point.
 
\item
The cluster problem is now set-up by calculating the 
cluster-excluded Green function ${\cal G}(K,\omega)$ as

\begin{equation} \label{eqn:clusterexG}
{\cal G}(K,\omega)
=
(\omega - \Gamma_o (K,\omega) - \overline{\epsilon}(K))^{-1}.
\end{equation}

Since the cluster problem is solved in real space, we then Fourier transform ${\cal G}$(K,$\omega$):
${\cal G}_{n,m} = \sum_{K} {\cal G}(K)\exp(i K\cdot(r_n-r_m))$.

\item
Next, we solve the cluster problem using, e.g., a MC simulation.  Here, we stochastically
generate random configurations of the disorder potential $V$.  For each disordered configuration, 
we use the Dyson equation to calculate the new fully dressed cluster Green function 
\begin{equation}
G^c(V) = ({\cal G}^{-1} - V)^{-1}.
\end{equation}
This is Fourier transformed to $G^c(K,K,\omega)$ to obtain the cluster density of states 
$\rho^c(K,\omega)=-\frac {1}{\pi} \textnormal{Im} G^c(K,K,\omega)$.  The typical cluster density 
of states is then calculated via geometric averaging using Eq.~\ref{eq:geometric_rho}. Then,
we calculate the disorder averaged, typical cluster Green function $G_{typ}^c(K,\omega)$ 
via Hilbert transform using Eq.~\ref{eq:Hilbert}.

\item
With the cluster problem solved, we use the typical cluster Green function $G_{typ}^c(K,\omega)$, 
to calculate the coarse-grained cluster Green function $\overline{G} (K,\omega)$ as

\begin{widetext}
\begin{equation} \label{eqn:coarsegrain}
\overline{G} (K,\omega)  
=
\frac{N_c}{N} \sum_{\tilde{k}} \frac{1}{(G^c_{typ} (K,\omega))^{-1} + \Gamma_o (K,\omega) - 
 \epsilon (K +\tilde{k}) + \overline{\epsilon}(K) + \mu}.
\end{equation}
\end{widetext}

\item
Finally, we calculate the new hybridization function using linear mixing

\begin{equation} \label{eqn:newhybrid}
\Gamma_n (K,\omega)
=
\Gamma_o (K,\omega) + \xi [(G^c_{typ} (K,\omega))^{-1} - (\overline{G} (K,\omega))^{-1}]
\end{equation}
where the subscripts $n$ and $o$ denote new and old, respectively. The mixing parameter 
$\xi > 0 $ controls the ratio of the new and old $\Gamma(K,\omega)$ entering the next iteration. 
For very small $\xi$, convergence may be slowed down unnecessarily, while for very 
large $\xi$, oscillations about the self-consistent solution may occur. Instead of 
linear mixing, the convergence of the computations can be improved by using the 
Broyden method~\cite{PhysRevB.38.12807}. 

\item
We repeat this procedure until the hybridization function converges to the desired 
accuracy, $\Gamma_o(K,\omega)=\Gamma_n(K,\omega)$. When this happens, the Green functions 
are also converged, $\overline{G}(K,\omega)= G^c_{typ}(K,\omega)$ within the computational error. 
\end{enumerate}

We note that instead of using the hybridization function $\Gamma(K,\omega)$ 
in the self-consistency, one can also use the 
self-energy $\Sigma(K,\omega)$. Both procedures should lead to the same solution 
since they are related by Eq.~\ref{eqn:oldhybrid}.

This method is causal, provided that the cluster solver produces 
a causal result. The argument is the same as that for the DCA\cite{PhysRevB.61.12739} 
and will not be repeated here.

\section{Results and discussion}
\label{sec:results}
We apply the CTMT scheme to one- and two-dimensional disordered systems describe by the Hamiltonian in Eq.~\ref{eqn:1}.
\begin{figure}[b]
\begin{center}
 \includegraphics[trim = 0mm 35mm 5mm 5mm,width=1\columnwidth,clip=true]{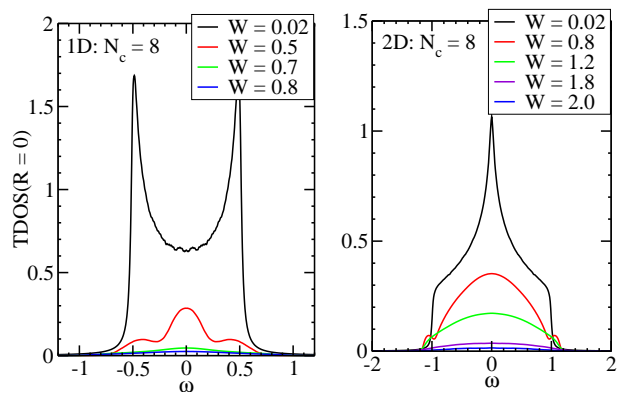}
\caption{(Color online) The local typical density of states, TDOS(R=0), as a function of 
frequency for various disorder strengths, $W$, for 
clusters of size N$_c$ = 8 for one- (left panel) and  two-(right panel) dimensional systems. Observe that in both cases, the TDOS gradually 
decreases with increasing $W$. The value of $W$ where the TDOS vanishes reveals the critical disorder strength $W_c$. 
Hence the TDOS behaves as an order parameter
for the localization transition. $\textnormal{Im}$ $\Gamma(R=0,\omega)$ 
(not shown) vanishes at the same critical disorder strength as the TDOS. }
\label{Tdos-w-Nc1-2D}
\end{center}
\end{figure}
Fig.~\ref{Tdos-w-Nc1-2D} shows the local typical density of states (TDOS(R=0)) 
for a cluster of size $N_c$ = 8 at 
various disorder strengths. 
For both one- and two-dimensional systems, the TDOS systematically goes 
to zero as the disorder strength is gradually increased. 
In 1D, the TDOS is practically zero for all frequencies at a critical disorder strength $W_c\approx 0.8$;
whereas in 2D, $W_c\approx 2.3$. Above the critical disorder strength, the electrons are localized. 
The TDOS calculated in our CTMT scheme indeed provides key information about the Anderson localization transition.

\begin{figure}
\begin{center}
 \includegraphics[trim = 0mm 0mm 0mm 0mm,width=1\columnwidth,clip=true]{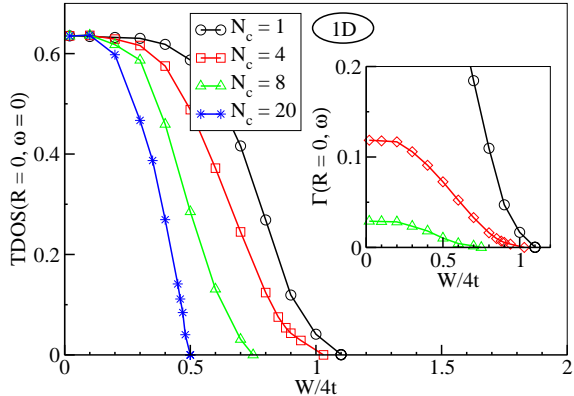}
\caption{(Color online) The local typical density of states at the band center, TDOS(R = 0, $\omega=0$), 
for one-dimension as a function of the disorder strength, W/4$t$, for various cluster sizes. The critical 
disorder strength $W_c$ decreases as the cluster size $N_c$ increases. The inset displays the local hybridization 
rate at the band center $\Gamma(R=0, \omega = 0)$ vs.\ W/4$t$. Note that $\Gamma(R=0, \omega=0)$ and 
TDOS(R=0, $\omega=0$) go to zero at the same critical disorder strength.}
\label{fig:TDOS-w-1D}
\end{center}
\end{figure}

According to the one-parameter scaling theory,~\cite{PhysRevLett.42.673,PhysRevLett.47.1546,PhysRev.109.1492,
RevModPhys.57.287} in 1D an arbitrary weak disorder strength localizes the electrons;
whereas in 2D, the system is also always localized, but with the difference
that the conductivity only decreases logarithmically with disorder strength.  

\begin{figure}
\begin{center}
 \includegraphics[trim = 0mm 0mm 0mm 0mm,width=1\columnwidth,clip=true]{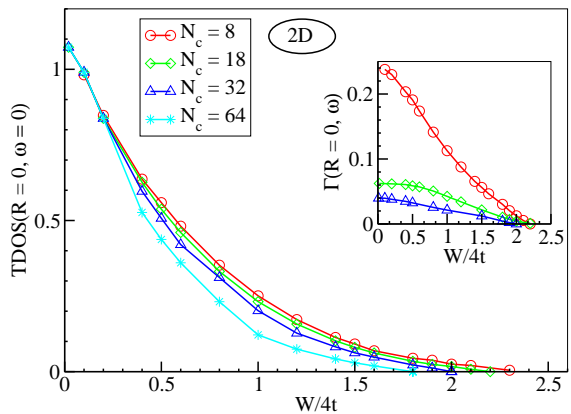}
\caption{(Color online) The local typical density of states at the band center, TDOS(R = 0, $\omega=0$), 
for a two-dimensional system as a function of disorder strength, W/4$t$, for various cluster sizes. 
The critical disorder strength $W_c$ decreases as cluster size $N_c$ increases.}
\label{fig:TDOS-w-2D}
\end{center}
\end{figure}

As we discussed in preceding sections, mean field theories, such as the single-site CPA, or the
DCA, cannot capture the localization transition due to the use of an algebraic averaging
scheme in their self-consistency. 
The single-site TMT, on the other hand, is able to qualitatively 
describe the localization transition in one, two and three dimensions,~\cite{PhysRevB.76.045105} 
but with critical disorder strength different from the exact values, i.e., in 1D and 2D, $W_c=0$ and in 
3D $W_c=2.1$ (in units where $4 t=1$).~\cite{PhysRevLett.82.382,PhysRevB.76.045105,Vlad2003} This is not 
surprising as the local TMT completely neglects non-local inter-site correlations.
The CTMT proposed in this paper gives a better qualitative and quantitative mean field theory for studying 
disordered systems in one and two-dimensions. We expect that as the cluster sizes increases, the critical 
disorder strength will systematically converge to the exact value of zero in the thermodynamic limit.
This is demonstrated in Figs.~\ref{fig:TDOS-w-1D} and \ref{fig:TDOS-w-2D}. 

Figure~\ref{fig:TDOS-w-1D} shows
the local typical density of states at the band center, TDOS(R = 0, $\omega=0$), for a one-dimensional 
system as a function of disorder strength, W. By increasing the 
cluster size the critical disorder strength decreases and eventually will go to zero for a
reasonably large $N_c$.
Similarly, Fig.~\ref{fig:TDOS-w-2D} displays the local typical density of states at
the band center for a two-dimensional system. The critical 
disorder strength also decreases as $N_c$ increase. However, 
the decrease of $W_c$ as a function of $N_c$ is slower than that of the 1D case. 
This implies that the scaling form of $W_c$ vs $N_c$ (or $L_c$) for 1D and 2D systems is different.
The former follows a power-law  whereas the latter has a logarithmic form. This is consistent with the 
one-parameter scaling theory~\cite{PhysRevLett.42.673,PhysRevLett.47.1546,PhysRev.109.1492,
RevModPhys.57.287} which shows that a 1D system is strongly localized, while 2D is 
the lower critical dimension of the Anderson
localization transition. 
The insets of Figs.~\ref{fig:TDOS-w-1D} and \ref{fig:TDOS-w-2D} show that the local hybridization rate 
at the band center, $\Gamma(R=0, \omega=0)$, goes to zero at the same value of $W$ than the TDOS.

\begin{figure}
\begin{center}
 \includegraphics[trim = 10mm 10mm 5mm 5mm,width=1\columnwidth,clip=true]{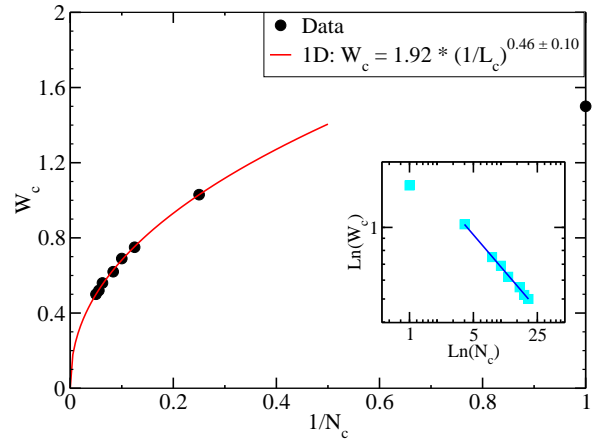}
\caption{(Color online) The critical disorder strengths $W_c$ for various cluster sizes 
$N_c$ ($=L_c$) in a 1D disordered system.  The data can be fitted with a scaling ansatz: 
$\xi = L_c \sim W_c^{-\nu}$  or $W_c \sim (1/L_c)^{1/\nu}$, with $1/\nu$ $=$ 0.46 $\pm$ 0.10 
in agreement with previous numerical results of $1/\nu$ $=$ 0.5\cite{JPSJ.81.104707}. 
The inset shows the power law behavior in a log-log plot.
}
\label{fig:Vc-Nc-1D}
\end{center}
\end{figure}

\begin{figure}[t!]
\begin{center}
  \includegraphics[trim = 0mm 0mm 0mm 0mm,width=1.0\columnwidth,clip=true]{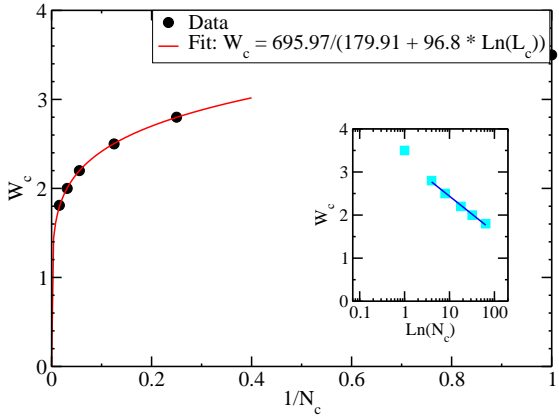}
 \caption{(Color online) The critical disorder strength $W_c$ for various cluster sizes 
$N_c=L^2_c$ in a 2D disordered system.  The data can be fit with a scaling ansatz: 
$W_c = 695.97/(179.91 + 96.8 \times \ln (L_c))$
in agreement with the results of \citet{PhysRevLett.47.1546}. 
The inset shows the logarithmic behavior in a semi-log plot.}
 \label{fig:Vc-Nc-2D}
\end{center}
 \end{figure}

Figs.~\ref{fig:Vc-Nc-1D} and \ref{fig:Vc-Nc-2D}  address the scaling of $W_c$ vs $N_c$.  
In 1D, the scaling ansatz is $\xi = L_c$ $\sim$ $W_c^{-\nu}$, where $\xi$ is the localization length, and
$\nu$ the critical exponent. As shown in Fig.~\ref{fig:Vc-Nc-1D}, 
the power law behavior is nicely captured by our data with
$1/\nu = 0.46 \pm 0.10$ in basic agreement with previous numerical results which reported a value of 
$1/\nu$ = 0.5,~\cite{JPSJ.81.104707} and also with 
the one-parameter scaling theory.~\cite{PhysRevLett.42.673} 
The inset of Fig.~\ref{fig:Vc-Nc-1D} is a log-log plot of $W_c$ vs $N_c$ 
where the power-law behavior $W_c \sim (1/L_c)^{1/\nu}$
is seen as a straight line.

Figure~\ref{fig:Vc-Nc-2D} displays the scaling of $W_c$ vs $L_c$ ($N_c=L^2_c$) in a 2D system.
Our data can be fit with a logarithmic function, 
$W_c = 695.97/(179.91 + 96.8 \times \ln (L_c))$. Therefore, the critical 
disorder strength  decreases slowly with cluster size and only 
for very large clusters the system is completely localized. This behavior agrees
with the one-parameter scaling theory,~\cite{PhysRevLett.42.673}
as 2D is generally believed to be the lower critical dimension 
for the Anderson localization.~\cite{PhysRevLett.47.882} 
The inset of Fig.~\ref{fig:Vc-Nc-2D} is a semi-log plot of $W_c$ vs $\ln$ $N_c$ 
where the logarithmic behavior is seen as a straight line. 
We note that our data fit nicely to an exponent form: 
$L_c = \exp(4.73/V_c^{0.98})/V_c^{0.98}$ in basic agreement with 
the results of \citet{PhysRevLett.47.1546}.

The present study clearly presents the essence of the effective typical medium theory, and 
the need to systematically go beyond single site approximations. 
Our formalism can easily be extended to study other physical phenomenon in lower 
dimensions. For example, in 1D systems, one can easily extend this approach to study super 
diffusion,\cite{PhysRevB.40.10999} which may not be possible in the local TMT approach, 
since the resonant states leading to super diffusion are non-local.
This present formalism can also be extended to study off-diagonal 
disorder in delocalization/localization processes in low dimensions.\cite{PhysRevB.4.2412,Eilmes200146,PSSB:PSSB205}

\begin{figure}[t]
 \includegraphics[trim = 0mm 0mm 0mm 0mm,width=1\columnwidth,clip=true]{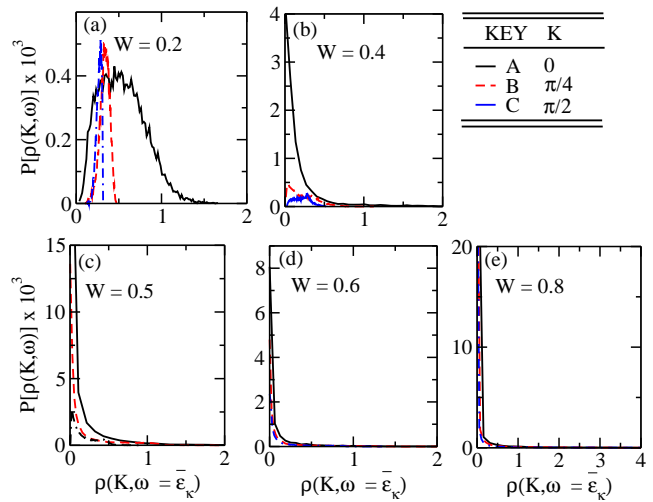}
\caption{(Color Online). The evolution of the  
probability distribution (PDF[$\rho(K,\omega=\bar{\epsilon}_K)$]) at 
different cluster cells with increasing $W$ for 1D system, the cluster size is N${}_c=8$. 
The labels A--C correspond to three cluster momenta. 
At very small disorder strength $W\approx0.4,0.5$, (cf. Fig.~\ref{Fig7:histrogram_1D}(b),(c)), 
all the cells show log-normal distribution.    
} 
\label{Fig7:histrogram_1D}
\end{figure}

\begin{figure}[h!]
 \includegraphics[trim = 0mm 0mm 0mm 0mm,width=1\columnwidth,clip=true]{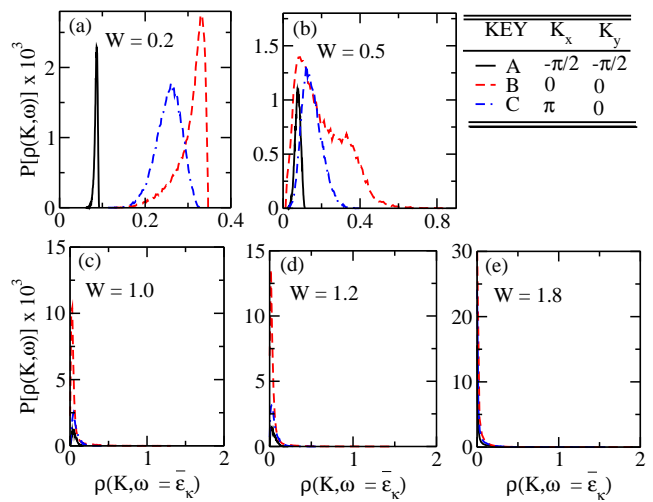}
\caption{(Color Online). The evolution of the  
probability distribution (PDF[$\rho(K,\omega=\bar{\epsilon}_K)$]) at 
different cluster cells with increasing $W$ for 2D system, the cluster size is N${}_c=8$. 
The labels A--C correspond to three cluster momenta. 
At small disorder strength, the PDF of some cells still remain Gaussian (cf. Fig.~\ref{Fig7:histrogram_1D}(a)-(b)). 
While as the disorder strength increases, (cf. Fig.~\ref{Fig7:histrogram_1D}(c),(d)), 
all the cells show log-normal distribution.     
} 
\label{Fig8:histrogram_2D}
\end{figure}

As explained in the previous sections, when system becomes localized, the PDFs of the density 
of states change from Gaussian distribution (where all states are metallic, 
the PDF is symmetric with the shape of TDOS the same as the ADOS) to a very asymmetric distribution 
with long tails (where all states are localized, the LDOS strongly fluctuating at all sites, and the TDOS is very different from the ADOS). Utilizing large-scale exact 
diagonalization calculations, 
Schubert \etal~\cite{PhysRevB.81.155106} have demonstrated that the PDFs close to the 
Anderson transition in 2D and 3D systems are log-normal.  
Here, as a proof of principle, we perform similar calculations. 
Within the CTMT scheme, we obtain the PDFs of the momentum-resolved 
DOS $\rho(K,\omega=\bar{\epsilon}_K)$ at different momenta cells centered at cluster momentum point $K$ and at the
averaged energy $\omega=\bar{\epsilon}_K$ of the cell $K$.
When sampled over large enough number of disorder configurations in our 
MC procedure, we indeed find as the disorder strength $W$ is increasing, 
the PDF[$\rho(K,\omega=\bar{\epsilon}_K)$] (shown in Figs.~\ref{Fig7:histrogram_1D} and \ref{Fig8:histrogram_2D} 
for 1D and 2D systems, respectively) develop log-normal distributions, 
consistent with the observation in Ref.~\onlinecite{PhysRevB.81.155106}. 
Moreover, we find for 1D system, the log-normal distribution happens for 
all the momentum cells at a very small disorder strength ($W=0.4,0.5$) for $Nc=8$ cluster; 
and in 2D, the same log-normal distribution for all the cells will only happen at a 
larger disorder strength ($W=1.0,1.2$). This observation is consistent with the fact that 
1D systems is much easier to localize as compared to 2D systems.
%


We also applied this approach to study three-dimensional systems. Unfortunately, our formalism, when applied
with modest cluster sizes, is unable to fully capture the localization transition in 3D.  First, we observe an 
unsystematic behavior of our formalism in the sense that for small cluster size, for e.g., N$_c$=1, 4 and 8, 
the critical disorder strength  $W_c$, where all states become localized, is underestimated,  while for larger 
cluster sizes, e.g., N$_c$=24 and 38, it overestimates the critical disorder strength.  Of course, in the 
limit of large N$_c$, the CTMT recovers the exact critical behavior around $W_c$.  Second, for modest cluster
sizes, while the TDOS becomes small as $W$ increases, its width increases monotonically with disorder strength
until the critical value is reached.   However, in exact diagonalization calculations the width first increases
and then decreases with disorder strength,\cite{PhysRevB.76.045105,Jung,Schubert} indicating that our current 
formalism fails to correctly capture the localization edge for modest cluster sizes. In addition, while 
the hybridization rates also become small, they do not all vanish at the critical disorder strength $W_c$. 
Rather, only the hybridizations corresponding to states near the top and bottom of the bands vanish 
while the the states at the band center only vanish for values of the disorder strength much 
larger than $W_c$.  Apparently, while our current CTMT formalism is able to capture weak localization 
effects in lower dimensional systems, it is not able to capture the evolution of the localization edge which
characterizes the transition in three dimensions until the cluster sizes are very large. We are working to develop 
a fully causal formalism which is able to efficiently capture the localization transition in 3D.

\section{Conclusion}
\label{sec:summary}
We develop a cluster extension to the local typical medium theory via the dynamical cluster 
approximation for studying localization in low dimensional disordered electronic systems. The 
developed CTMT systematically incorporates nonlocal corrections  to capture quantum coherence. 
The formalism recovers the local TMT when the cluster size is $N_c=1$, and becomes exact as 
N$_c$ $\rightarrow$ $\infty$.  Such an approach opens a new avenue to study localization 
effects in lower dimensional model systems.

{\bf Acknowledgments.}
We thank K. M. Tam and S. X. Yang for useful discussions. 
Work at LSU is funded by the National Science Foundation LA-SiGMA award: EPS-1003897. Work at 
BNL is supported by the U.S. Department of Energy (DOE) under contract DE-AC02-98CH10886.
High performance computational resources are provided by Louisiana Optical Network Initiative 
(LONI), and HPC@LSU computing resources.
Work at FSU is supported by the National High Magnetic Field Laboratory and the NSF Grant No. DMR-1005751.


\begin{thebibliography}{54}
\expandafter\ifx\csname natexlab\endcsname\relax\def\natexlab#1{#1}\fi
\expandafter\ifx\csname bibnamefont\endcsname\relax
  \def\bibnamefont#1{#1}\fi
\expandafter\ifx\csname bibfnamefont\endcsname\relax
  \def\bibfnamefont#1{#1}\fi
\expandafter\ifx\csname citenamefont\endcsname\relax
  \def\citenamefont#1{#1}\fi
\expandafter\ifx\csname url\endcsname\relax
  \def\url#1{\texttt{#1}}\fi
\expandafter\ifx\csname urlprefix\endcsname\relax\def\urlprefix{URL }\fi
\providecommand{\bibinfo}[2]{#2}
\providecommand{\eprint}[2][]{\url{#2}}

\bibitem[{\citenamefont{Abrahams et~al.}(1979)\citenamefont{Abrahams, Anderson,
  Licciardello, and Ramakrishnan}}]{PhysRevLett.42.673}
\bibinfo{author}{\bibfnamefont{E.}~\bibnamefont{Abrahams}},
  \bibinfo{author}{\bibfnamefont{P.~W.} \bibnamefont{Anderson}},
  \bibinfo{author}{\bibfnamefont{D.~C.} \bibnamefont{Licciardello}},
  \bibnamefont{and} \bibinfo{author}{\bibfnamefont{T.~V.}
  \bibnamefont{Ramakrishnan}}, \bibinfo{journal}{Phys. Rev. Lett.}
  \textbf{\bibinfo{volume}{42}}, \bibinfo{pages}{673} (\bibinfo{year}{1979}).

\bibitem[{\citenamefont{MacKinnon and Kramer}(1981)}]{PhysRevLett.47.1546}
\bibinfo{author}{\bibfnamefont{A.}~\bibnamefont{MacKinnon}} \bibnamefont{and}
  \bibinfo{author}{\bibfnamefont{B.}~\bibnamefont{Kramer}},
  \bibinfo{journal}{Phys. Rev. Lett.} \textbf{\bibinfo{volume}{47}},
  \bibinfo{pages}{1546} (\bibinfo{year}{1981}).

\bibitem[{\citenamefont{Anderson}(1958)}]{PhysRev.109.1492}
\bibinfo{author}{\bibfnamefont{P.~W.} \bibnamefont{Anderson}},
  \bibinfo{journal}{Phys. Rev.} \textbf{\bibinfo{volume}{109}},
  \bibinfo{pages}{1492} (\bibinfo{year}{1958}).

\bibitem[{\citenamefont{Lagendijk et~al.}(2009)\citenamefont{Lagendijk, van
  Tiggelen, and Wiersma}}]{Lagendijk2009}
\bibinfo{author}{\bibfnamefont{A.}~\bibnamefont{Lagendijk}},
  \bibinfo{author}{\bibfnamefont{B.}~\bibnamefont{van Tiggelen}},
  \bibnamefont{and} \bibinfo{author}{\bibfnamefont{D.~S.}
  \bibnamefont{Wiersma}}, \bibinfo{journal}{Physics Today}
  \textbf{\bibinfo{volume}{62}}, \bibinfo{pages}{24} (\bibinfo{year}{2009}).

\bibitem[{\citenamefont{Lee and Ramakrishnan}(1985)}]{RevModPhys.57.287}
\bibinfo{author}{\bibfnamefont{P.~A.} \bibnamefont{Lee}} \bibnamefont{and}
  \bibinfo{author}{\bibfnamefont{T.~V.} \bibnamefont{Ramakrishnan}},
  \bibinfo{journal}{Rev. Mod. Phys.} \textbf{\bibinfo{volume}{57}},
  \bibinfo{pages}{287} (\bibinfo{year}{1985}).

\bibitem[{\citenamefont{Slevin and Ohtsuki}(1999)}]{PhysRevLett.82.382}
\bibinfo{author}{\bibfnamefont{K.}~\bibnamefont{Slevin}} \bibnamefont{and}
  \bibinfo{author}{\bibfnamefont{T.}~\bibnamefont{Ohtsuki}},
  \bibinfo{journal}{Phys. Rev. Lett.} \textbf{\bibinfo{volume}{82}},
  \bibinfo{pages}{382} (\bibinfo{year}{1999}).

\bibitem[{\citenamefont{Schubert et~al.}(2010)\citenamefont{Schubert, Schleede,
  Byczuk, Fehske, and Vollhardt}}]{PhysRevB.81.155106}
\bibinfo{author}{\bibfnamefont{G.}~\bibnamefont{Schubert}},
  \bibinfo{author}{\bibfnamefont{J.}~\bibnamefont{Schleede}},
  \bibinfo{author}{\bibfnamefont{K.}~\bibnamefont{Byczuk}},
  \bibinfo{author}{\bibfnamefont{H.}~\bibnamefont{Fehske}}, \bibnamefont{and}
  \bibinfo{author}{\bibfnamefont{D.}~\bibnamefont{Vollhardt}},
  \bibinfo{journal}{Phys. Rev. B} \textbf{\bibinfo{volume}{81}},
  \bibinfo{pages}{155106} (\bibinfo{year}{2010}).

\bibitem[{\citenamefont{Pichard and Sarma}(1981)}]{Pichard1981}
\bibinfo{author}{\bibfnamefont{J.~P.} \bibnamefont{Pichard}} \bibnamefont{and}
  \bibinfo{author}{\bibfnamefont{G.}~\bibnamefont{Sarma}}, \bibinfo{journal}{J.
  Phys. C} \textbf{\bibinfo{volume}{14}}, \bibinfo{pages}{L127}
  (\bibinfo{year}{1981}).

\bibitem[{\citenamefont{Dobrosavljevi\'{c}
  et~al.}(2003)\citenamefont{Dobrosavljevi\'{c}, Pastor, and
  Nikoli\'{c}}}]{Vlad2003}
\bibinfo{author}{\bibfnamefont{V.}~\bibnamefont{Dobrosavljevi\'{c}}},
  \bibinfo{author}{\bibfnamefont{A.~A.} \bibnamefont{Pastor}},
  \bibnamefont{and} \bibinfo{author}{\bibfnamefont{B.~K.}
  \bibnamefont{Nikoli\'{c}}}, \bibinfo{journal}{Europhysics Letters}
  \textbf{\bibinfo{volume}{62}}, \bibinfo{pages}{76} (\bibinfo{year}{2003}).

\bibitem[{\citenamefont{Elliott et~al.}(1974)\citenamefont{Elliott, Krumhansl,
  and Leath}}]{RevModPhys.46.465}
\bibinfo{author}{\bibfnamefont{R.~J.} \bibnamefont{Elliott}},
  \bibinfo{author}{\bibfnamefont{J.~A.} \bibnamefont{Krumhansl}},
  \bibnamefont{and} \bibinfo{author}{\bibfnamefont{P.~L.} \bibnamefont{Leath}},
  \bibinfo{journal}{Rev. Mod. Phys.} \textbf{\bibinfo{volume}{46}},
  \bibinfo{pages}{465} (\bibinfo{year}{1974}).

\bibitem[{\citenamefont{Song et~al.}(2007)\citenamefont{Song, Atkinson, and
  Wortis}}]{PhysRevB.76.045105}
\bibinfo{author}{\bibfnamefont{Y.}~\bibnamefont{Song}},
  \bibinfo{author}{\bibfnamefont{W.~A.} \bibnamefont{Atkinson}},
  \bibnamefont{and} \bibinfo{author}{\bibfnamefont{R.}~\bibnamefont{Wortis}},
  \bibinfo{journal}{Phys. Rev. B} \textbf{\bibinfo{volume}{76}},
  \bibinfo{pages}{045105} (\bibinfo{year}{2007}).

\bibitem[{\citenamefont{Bhatt and Johri}(2012)}]{Bhatt2012}
\bibinfo{author}{\bibfnamefont{R.~N.} \bibnamefont{Bhatt}} \bibnamefont{and}
  \bibinfo{author}{\bibfnamefont{S.}~\bibnamefont{Johri}},
  \bibinfo{journal}{Int. J. Mod. Phys.: Conf. Ser.}
  \textbf{\bibinfo{volume}{11}}, \bibinfo{pages}{79 } (\bibinfo{year}{2012}).

\bibitem[{\citenamefont{Gorkov et~al.}(1979)\citenamefont{Gorkov, Larkin, and
  Khmelnitskii}}]{Gorkov1979}
\bibinfo{author}{\bibfnamefont{L.~P.} \bibnamefont{Gorkov}},
  \bibinfo{author}{\bibfnamefont{A.~I.} \bibnamefont{Larkin}},
  \bibnamefont{and} \bibinfo{author}{\bibfnamefont{D.~E.}
  \bibnamefont{Khmelnitskii}}, \bibinfo{journal}{Sov. Phys. JETP Lett.}
  \textbf{\bibinfo{volume}{30}}, \bibinfo{pages}{228} (\bibinfo{year}{1979}).

\bibitem[{\citenamefont{Clark}(1967)}]{PhysRev.154.750}
\bibinfo{author}{\bibfnamefont{A.~H.} \bibnamefont{Clark}},
  \bibinfo{journal}{Phys. Rev.} \textbf{\bibinfo{volume}{154}},
  \bibinfo{pages}{750} (\bibinfo{year}{1967}).

\bibitem[{\citenamefont{Morgan and Walley}(1971)}]{Morgan1971}
\bibinfo{author}{\bibfnamefont{M.}~\bibnamefont{Morgan}} \bibnamefont{and}
  \bibinfo{author}{\bibfnamefont{P.~A.} \bibnamefont{Walley}},
  \bibinfo{journal}{Philos. Mag.} \textbf{\bibinfo{volume}{23}},
  \bibinfo{pages}{661} (\bibinfo{year}{1971}).

\bibitem[{\citenamefont{Kramer and MacKinnon}(1993)}]{Kramer1993}
\bibinfo{author}{\bibfnamefont{B.}~\bibnamefont{Kramer}} \bibnamefont{and}
  \bibinfo{author}{\bibfnamefont{A.}~\bibnamefont{MacKinnon}},
  \bibinfo{journal}{Reports on Progress in Physics}
  \textbf{\bibinfo{volume}{56}}, \bibinfo{pages}{1469} (\bibinfo{year}{1993}).

\bibitem[{\citenamefont{Sharvin and Sharvin}(1982)}]{Sharvin1982}
\bibinfo{author}{\bibfnamefont{D.~Y.} \bibnamefont{Sharvin}} \bibnamefont{and}
  \bibinfo{author}{\bibfnamefont{Y.~V.} \bibnamefont{Sharvin}},
  \bibinfo{journal}{Sov. Phys. JETP Lett.} \textbf{\bibinfo{volume}{34}},
  \bibinfo{pages}{272} (\bibinfo{year}{1982}).

\bibitem[{\citenamefont{Bergmann}(1984)}]{Bergmann1984}
\bibinfo{author}{\bibfnamefont{G.}~\bibnamefont{Bergmann}},
  \bibinfo{journal}{Phys. Rep.} \textbf{\bibinfo{volume}{107}},
  \bibinfo{pages}{1} (\bibinfo{year}{1984}).

\bibitem[{\citenamefont{Kravchenko et~al.}(1994)\citenamefont{Kravchenko,
  Kravchenko, Furneaux, Pudalov, and D'Iorio}}]{PhysRevB.50.8039}
\bibinfo{author}{\bibfnamefont{S.~V.} \bibnamefont{Kravchenko}},
  \bibinfo{author}{\bibfnamefont{G.~V.} \bibnamefont{Kravchenko}},
  \bibinfo{author}{\bibfnamefont{J.~E.} \bibnamefont{Furneaux}},
  \bibinfo{author}{\bibfnamefont{V.~M.} \bibnamefont{Pudalov}},
  \bibnamefont{and} \bibinfo{author}{\bibfnamefont{M.}~\bibnamefont{D'Iorio}},
  \bibinfo{journal}{Phys. Rev. B} \textbf{\bibinfo{volume}{50}},
  \bibinfo{pages}{8039} (\bibinfo{year}{1994}).

\bibitem[{\citenamefont{Kravchenko and Sarachik}(2004)}]{Kravchenko2004}
\bibinfo{author}{\bibfnamefont{S.~V.} \bibnamefont{Kravchenko}}
  \bibnamefont{and} \bibinfo{author}{\bibfnamefont{M.~P.}
  \bibnamefont{Sarachik}}, \bibinfo{journal}{Rep. Prog. Phys.}
  \textbf{\bibinfo{volume}{67}}, \bibinfo{pages}{1} (\bibinfo{year}{2004}).

\bibitem[{\citenamefont{Abrahams et~al.}(2001)\citenamefont{Abrahams,
  Kravchenko, and Sarachik}}]{RevModPhys.73.251}
\bibinfo{author}{\bibfnamefont{E.}~\bibnamefont{Abrahams}},
  \bibinfo{author}{\bibfnamefont{S.~V.} \bibnamefont{Kravchenko}},
  \bibnamefont{and} \bibinfo{author}{\bibfnamefont{M.~P.}
  \bibnamefont{Sarachik}}, \bibinfo{journal}{Rev. Mod. Phys.}
  \textbf{\bibinfo{volume}{73}}, \bibinfo{pages}{251} (\bibinfo{year}{2001}).

\bibitem[{\citenamefont{Billy et~al.}(2008)\citenamefont{Billy, Josse, Zuo,
  Bernard, Hambrecht, Lugan, Clement, Sanchez-Palencia, Bouyer, and
  Aspect}}]{Billy2008}
\bibinfo{author}{\bibfnamefont{J.}~\bibnamefont{Billy}},
  \bibinfo{author}{\bibfnamefont{V.}~\bibnamefont{Josse}},
  \bibinfo{author}{\bibfnamefont{Z.}~\bibnamefont{Zuo}},
  \bibinfo{author}{\bibfnamefont{A.}~\bibnamefont{Bernard}},
  \bibinfo{author}{\bibfnamefont{B.}~\bibnamefont{Hambrecht}},
  \bibinfo{author}{\bibfnamefont{P.}~\bibnamefont{Lugan}},
  \bibinfo{author}{\bibfnamefont{D.}~\bibnamefont{Clement}},
  \bibinfo{author}{\bibfnamefont{L.}~\bibnamefont{Sanchez-Palencia}},
  \bibinfo{author}{\bibfnamefont{P.}~\bibnamefont{Bouyer}}, \bibnamefont{and}
  \bibinfo{author}{\bibfnamefont{A.}~\bibnamefont{Aspect}},
  \bibinfo{journal}{Nature} \textbf{\bibinfo{volume}{453}},
  \bibinfo{pages}{891} (\bibinfo{year}{2008}).

\bibitem[{\citenamefont{Roati et~al.}(2008)\citenamefont{Roati, D'Errico,
  Fallani, Fattori, Fort, Zaccanti, Modugno, Modugno, and
  Inguscio}}]{Roati2008}
\bibinfo{author}{\bibfnamefont{G.}~\bibnamefont{Roati}},
  \bibinfo{author}{\bibfnamefont{C.}~\bibnamefont{D'Errico}},
  \bibinfo{author}{\bibfnamefont{L.}~\bibnamefont{Fallani}},
  \bibinfo{author}{\bibfnamefont{M.}~\bibnamefont{Fattori}},
  \bibinfo{author}{\bibfnamefont{C.}~\bibnamefont{Fort}},
  \bibinfo{author}{\bibfnamefont{M.}~\bibnamefont{Zaccanti}},
  \bibinfo{author}{\bibfnamefont{G.}~\bibnamefont{Modugno}},
  \bibinfo{author}{\bibfnamefont{M.}~\bibnamefont{Modugno}}, \bibnamefont{and}
  \bibinfo{author}{\bibfnamefont{M.}~\bibnamefont{Inguscio}},
  \bibinfo{journal}{Nature} \textbf{\bibinfo{volume}{453}},
  \bibinfo{pages}{895} (\bibinfo{year}{2008}).

\bibitem[{\citenamefont{Chab\'e et~al.}(2008)\citenamefont{Chab\'e, Lemari\'e,
  Gr\'emaud, Delande, Szriftgiser, and Garreau}}]{PhysRevLett.101.255702}
\bibinfo{author}{\bibfnamefont{J.}~\bibnamefont{Chab\'e}},
  \bibinfo{author}{\bibfnamefont{G.}~\bibnamefont{Lemari\'e}},
  \bibinfo{author}{\bibfnamefont{B.}~\bibnamefont{Gr\'emaud}},
  \bibinfo{author}{\bibfnamefont{D.}~\bibnamefont{Delande}},
  \bibinfo{author}{\bibfnamefont{P.}~\bibnamefont{Szriftgiser}},
  \bibnamefont{and} \bibinfo{author}{\bibfnamefont{J.~C.}
  \bibnamefont{Garreau}}, \bibinfo{journal}{Phys. Rev. Lett.}
  \textbf{\bibinfo{volume}{101}}, \bibinfo{pages}{255702}
  (\bibinfo{year}{2008}).

\bibitem[{\citenamefont{Jarrell and Krishnamurthy}(2001)}]{PhysRevB.63.125102}
\bibinfo{author}{\bibfnamefont{M.}~\bibnamefont{Jarrell}} \bibnamefont{and}
  \bibinfo{author}{\bibfnamefont{H.~R.} \bibnamefont{Krishnamurthy}},
  \bibinfo{journal}{Phys. Rev. B} \textbf{\bibinfo{volume}{63}},
  \bibinfo{pages}{125102} (\bibinfo{year}{2001}).

\bibitem[{\citenamefont{Maier et~al.}(2005)\citenamefont{Maier, Jarrell,
  Pruschke, and Hettler}}]{RevModPhys.77.1027}
\bibinfo{author}{\bibfnamefont{T.}~\bibnamefont{Maier}},
  \bibinfo{author}{\bibfnamefont{M.}~\bibnamefont{Jarrell}},
  \bibinfo{author}{\bibfnamefont{T.}~\bibnamefont{Pruschke}}, \bibnamefont{and}
  \bibinfo{author}{\bibfnamefont{M.~H.} \bibnamefont{Hettler}},
  \bibinfo{journal}{Rev. Mod. Phys.} \textbf{\bibinfo{volume}{77}},
  \bibinfo{pages}{1027} (\bibinfo{year}{2005}).

\bibitem[{\citenamefont{Hettler et~al.}(2000)\citenamefont{Hettler, Mukherjee,
  Jarrell, and Krishnamurthy}}]{PhysRevB.61.12739}
\bibinfo{author}{\bibfnamefont{M.~H.} \bibnamefont{Hettler}},
  \bibinfo{author}{\bibfnamefont{M.}~\bibnamefont{Mukherjee}},
  \bibinfo{author}{\bibfnamefont{M.}~\bibnamefont{Jarrell}}, \bibnamefont{and}
  \bibinfo{author}{\bibfnamefont{H.~R.} \bibnamefont{Krishnamurthy}},
  \bibinfo{journal}{Phys. Rev. B} \textbf{\bibinfo{volume}{61}},
  \bibinfo{pages}{12739} (\bibinfo{year}{2000}).

\bibitem[{mvp(2002)}]{mvp1}
\bibinfo{journal}{The most probable value of a random quantity is the mode,
  which is the value for which its PDF becomes maximal. A property X of a given
  system is self-averaging if 'most' realizations of the randomness in the
  thermodynamic limit have the same value of X. The Anderson localization does
  not have this property. Close to the critical point, physical observables are
  not Gaussian and generally have log-normal behavior. For discussions, see A.
  Aharony and A. B. Harris, Phys. Rev. Lett. \textbf{77}, 3700 -- 3703 (1996);
  S. Wiseman and E. Domany, Phys. Rev. E \textbf{52}, 3469 -- 3484 (1995), E.
  Orlandini, M .C. Tesi, and S. G. Whittington, J. Phys. A: Math. Gen.
  \textbf{35}, 4219 -- 4227.}  (\bibinfo{year}{2002}).

\bibitem[{\citenamefont{Mirlin and Fyodorov}(1994)}]{PhysRevLett.72.526}
\bibinfo{author}{\bibfnamefont{A.~D.} \bibnamefont{Mirlin}} \bibnamefont{and}
  \bibinfo{author}{\bibfnamefont{Y.~V.} \bibnamefont{Fyodorov}},
  \bibinfo{journal}{Phys. Rev. Lett.} \textbf{\bibinfo{volume}{72}},
  \bibinfo{pages}{526} (\bibinfo{year}{1994}).

\bibitem[{\citenamefont{Janssen}(1994)}]{Janssen1994}
\bibinfo{author}{\bibfnamefont{M.}~\bibnamefont{Janssen}},
  \bibinfo{journal}{Int. J. Mod. Phys. B} \textbf{\bibinfo{volume}{8}},
  \bibinfo{pages}{943} (\bibinfo{year}{1994}).

\bibitem[{\citenamefont{Byczuk et~al.}(2005)\citenamefont{Byczuk, Hofstetter,
  and Vollhardt}}]{PhysRevLett.94.056404}
\bibinfo{author}{\bibfnamefont{K.}~\bibnamefont{Byczuk}},
  \bibinfo{author}{\bibfnamefont{W.}~\bibnamefont{Hofstetter}},
  \bibnamefont{and}
  \bibinfo{author}{\bibfnamefont{D.}~\bibnamefont{Vollhardt}},
  \bibinfo{journal}{Phys. Rev. Lett.} \textbf{\bibinfo{volume}{94}},
  \bibinfo{pages}{056404} (\bibinfo{year}{2005}).

\bibitem[{\citenamefont{Thouless}(1974)}]{Thouless1974}
\bibinfo{author}{\bibfnamefont{D.~J.} \bibnamefont{Thouless}},
  \bibinfo{journal}{Phys. Reports} \textbf{\bibinfo{volume}{13}},
  \bibinfo{pages}{93--142} (\bibinfo{year}{1974}).

\bibitem[{\citenamefont{Thouless}(1970)}]{Thouless1970}
\bibinfo{author}{\bibfnamefont{D.~J.} \bibnamefont{Thouless}},
  \bibinfo{journal}{J. Phys. C: Solid State Phys.} \textbf{\bibinfo{volume}{3}},
  \bibinfo{pages}{1559} (\bibinfo{year}{1970}).

\bibitem[{\citenamefont{Janssen}(1998)}]{Janssen1998}
\bibinfo{author}{\bibfnamefont{M.}~\bibnamefont{Janssen}},
  \bibinfo{journal}{Phys. Rep.} \textbf{\bibinfo{volume}{295}},
  \bibinfo{pages}{1} (\bibinfo{year}{1998}).

\bibitem[{\citenamefont{Brndiar and Marko\ifmmode~\check{s}\else
  \v{s}\fi{}}(2006)}]{PhysRevB.74.153103}
\bibinfo{author}{\bibfnamefont{J.}~\bibnamefont{Brndiar}} \bibnamefont{and}
  \bibinfo{author}{\bibfnamefont{P.}~\bibnamefont{Marko\ifmmode~\check{s}\else
  \v{s}\fi{}}}, \bibinfo{journal}{Phys. Rev. B} \textbf{\bibinfo{volume}{74}},
  \bibinfo{pages}{153103} (\bibinfo{year}{2006}).

\bibitem[{\citenamefont{Crow and Shimizu}(1988)}]{Crow1988}
\bibinfo{editor}{\bibfnamefont{E.}~\bibnamefont{Crow}} \bibnamefont{and}
  \bibinfo{editor}{\bibfnamefont{K.}~\bibnamefont{Shimizu}}, eds.,
  \emph{\bibinfo{title}{Log-Normal Distribution--Theory and Applications}}
  (\bibinfo{publisher}{Marcel Dekker, Inc., New York}, \bibinfo{year}{1988}).

\bibitem[{ago(1992)}]{agonis}
\emph{\bibinfo{title}{The problems encountered in early attempts to formulate
  cluster corrections to the DMFA are the same as those encountered in the
  coherent potential approximation (CPA) for disordered systems. For a detailed
  discussion of earlier work on the inclusion of non-local corrections to the
  CPA see {\em{A. Gonis, Green functions for ordered and disordered systems}},
  in the series {\em{Studies in Mathematical Physics. Eds. E.~van~Groesen and
  E.~M.~DeJager}}}} (\bibinfo{publisher}{North Holland, Amsterdam},
  \bibinfo{year}{1992}).

\bibitem[{\citenamefont{Ishii}(1973)}]{Ishii1973}
\bibinfo{author}{\bibfnamefont{K.}~\bibnamefont{Ishii}},
  \bibinfo{journal}{Prog. Theor. Phys. Suppl.} \textbf{\bibinfo{volume}{53}},
  \bibinfo{pages}{77} (\bibinfo{year}{1973}).

\bibitem[{\citenamefont{Vollhardt and W\"olfle}(1980)}]{PhysRevLett.45.842}
\bibinfo{author}{\bibfnamefont{D.}~\bibnamefont{Vollhardt}} \bibnamefont{and}
  \bibinfo{author}{\bibfnamefont{P.}~\bibnamefont{W\"olfle}},
  \bibinfo{journal}{Phys. Rev. Lett.} \textbf{\bibinfo{volume}{45}},
  \bibinfo{pages}{842} (\bibinfo{year}{1980}).

\bibitem[{\citenamefont{Thouless}(2010)}]{Thouless2010}
\bibinfo{author}{\bibfnamefont{D.}~\bibnamefont{Thouless}},
  \bibinfo{journal}{Inter. J. Moder. Phys. B} \textbf{\bibinfo{volume}{24}},
  \bibinfo{pages}{1507} (\bibinfo{year}{2010}).

\bibitem[{\citenamefont{MacKinnon and Kramer}(1983)}]{MacKinnon1983}
\bibinfo{author}{\bibfnamefont{A.}~\bibnamefont{MacKinnon}} \bibnamefont{and}
  \bibinfo{author}{\bibfnamefont{B.}~\bibnamefont{Kramer}},
  \bibinfo{journal}{Z. Physik B Condensed Matter}
  \textbf{\bibinfo{volume}{53}}, \bibinfo{pages}{1} (\bibinfo{year}{1983}).

\bibitem[{\citenamefont{Wischmann and M\"{u}ller-Hartmann}(1990)}]{Wischmann}
\bibinfo{author}{\bibfnamefont{B.}~\bibnamefont{Wischmann}} \bibnamefont{and}
  \bibinfo{author}{\bibfnamefont{E.}~\bibnamefont{M\"{u}ller-Hartmann}},
  \bibinfo{journal}{Z. Physik B Condensed Matter}
  \textbf{\bibinfo{volume}{79}}, \bibinfo{pages}{91} (\bibinfo{year}{1990}).

\bibitem[{\citenamefont{Soven}(1967)}]{PhysRev.156.809}
\bibinfo{author}{\bibfnamefont{P.}~\bibnamefont{Soven}},
  \bibinfo{journal}{Phys. Rev.} \textbf{\bibinfo{volume}{156}},
  \bibinfo{pages}{809} (\bibinfo{year}{1967}).

\bibitem[{\citenamefont{Gross et~al.}(1991)\citenamefont{Gross, Runge, and
  Heinonen}}]{Gross1991}
\bibinfo{author}{\bibfnamefont{E.}~\bibnamefont{Gross}},
  \bibinfo{author}{\bibfnamefont{E.}~\bibnamefont{Runge}}, \bibnamefont{and}
  \bibinfo{author}{\bibfnamefont{O.}~\bibnamefont{Heinonen}},
  \emph{\bibinfo{title}{Many-Particle Theory}} (\bibinfo{publisher}{A. Hilger,
  Bristol}, \bibinfo{year}{1991}), ISBN \bibinfo{isbn}{9780750301558}.

\bibitem[{\citenamefont{Fetter and Walecka}(1971)}]{fetter1971quantum}
\bibinfo{author}{\bibfnamefont{A.}~\bibnamefont{Fetter}} \bibnamefont{and}
  \bibinfo{author}{\bibfnamefont{J.}~\bibnamefont{Walecka}},
  \emph{\bibinfo{title}{Quantum Theory of Many-particle Systems}}, Dover books
  on physics (\bibinfo{publisher}{Dover Publishers Incorporated},
  \bibinfo{year}{1971}), ISBN \bibinfo{isbn}{9780486428277}.

\bibitem[{\citenamefont{Johnson}(1988)}]{PhysRevB.38.12807}
\bibinfo{author}{\bibfnamefont{D.~D.} \bibnamefont{Johnson}},
  \bibinfo{journal}{Phys. Rev. B} \textbf{\bibinfo{volume}{38}},
  \bibinfo{pages}{12807} (\bibinfo{year}{1988}).

\bibitem[{\citenamefont{Yakubo and Mizutaka}(2012)}]{JPSJ.81.104707}
\bibinfo{author}{\bibfnamefont{K.}~\bibnamefont{Yakubo}} \bibnamefont{and}
  \bibinfo{author}{\bibfnamefont{S.}~\bibnamefont{Mizutaka}},
  \bibinfo{journal}{J. Phys. Soc. Jpn} \textbf{\bibinfo{volume}{81}},
  \bibinfo{pages}{104707} (\bibinfo{year}{2012}).

\bibitem[{\citenamefont{Lee and Fisher}(1981)}]{PhysRevLett.47.882}
\bibinfo{author}{\bibfnamefont{P.~A.} \bibnamefont{Lee}} \bibnamefont{and}
  \bibinfo{author}{\bibfnamefont{D.~S.} \bibnamefont{Fisher}},
  \bibinfo{journal}{Phys. Rev. Lett.} \textbf{\bibinfo{volume}{47}},
  \bibinfo{pages}{882} (\bibinfo{year}{1981}).

\bibitem[{\citenamefont{Dunlap et~al.}(1989)\citenamefont{Dunlap, Kundu, and
  Phillips}}]{PhysRevB.40.10999}
\bibinfo{author}{\bibfnamefont{D.~H.} \bibnamefont{Dunlap}},
  \bibinfo{author}{\bibfnamefont{K.}~\bibnamefont{Kundu}}, \bibnamefont{and}
  \bibinfo{author}{\bibfnamefont{P.}~\bibnamefont{Phillips}},
  \bibinfo{journal}{Phys. Rev. B} \textbf{\bibinfo{volume}{40}},
  \bibinfo{pages}{10999} (\bibinfo{year}{1989}).

\bibitem[{\citenamefont{Blackman et~al.}(1971)\citenamefont{Blackman,
  Esterling, and Berk}}]{PhysRevB.4.2412}
\bibinfo{author}{\bibfnamefont{J.~A.} \bibnamefont{Blackman}},
  \bibinfo{author}{\bibfnamefont{D.~M.} \bibnamefont{Esterling}},
  \bibnamefont{and} \bibinfo{author}{\bibfnamefont{N.~F.} \bibnamefont{Berk}},
  \bibinfo{journal}{Phys. Rev. B} \textbf{\bibinfo{volume}{4}},
  \bibinfo{pages}{2412} (\bibinfo{year}{1971}).

\bibitem[{\citenamefont{Eilmes et~al.}(2001)\citenamefont{Eilmes, R\"{o}mer,
  and Schreiber}}]{Eilmes200146}
\bibinfo{author}{\bibfnamefont{A.}~\bibnamefont{Eilmes}},
  \bibinfo{author}{\bibfnamefont{R.~A.} \bibnamefont{R\"{o}mer}},
  \bibnamefont{and}
  \bibinfo{author}{\bibfnamefont{M.}~\bibnamefont{Schreiber}},
  \bibinfo{journal}{Physica B: Condensed Matter}
  \textbf{\bibinfo{volume}{296}}, \bibinfo{pages}{46 } (\bibinfo{year}{2001}).

\bibitem[{\citenamefont{Biswas et~al.}(2000)\citenamefont{Biswas, Cain,
  R\"{o}mer, and Schreiber}}]{PSSB:PSSB205}
\bibinfo{author}{\bibfnamefont{P.}~\bibnamefont{Biswas}},
  \bibinfo{author}{\bibfnamefont{P.}~\bibnamefont{Cain}},
  \bibinfo{author}{\bibfnamefont{R.}~\bibnamefont{R\"{o}mer}},
  \bibnamefont{and}
  \bibinfo{author}{\bibfnamefont{M.}~\bibnamefont{Schreiber}},
  \bibinfo{journal}{physica status solidi (b)} \textbf{\bibinfo{volume}{218}},
  \bibinfo{pages}{205} (\bibinfo{year}{2000}).

\bibitem[{\citenamefont{Jung et~al.}(2012)\citenamefont{Jung, Czycholl, and
  Kettemann}}]{Jung}
\bibinfo{author}{\bibfnamefont{D.}~\bibnamefont{Jung}},
  \bibinfo{author}{\bibfnamefont{G.}~\bibnamefont{Czycholl}}, \bibnamefont{and}
  \bibinfo{author}{\bibfnamefont{S.}~\bibnamefont{Kettemann}},
  \bibinfo{journal}{Inter. J. Moder. Phys.: Conference Series}
  \textbf{\bibinfo{volume}{11}}, \bibinfo{pages}{108 } (\bibinfo{year}{2012}).

\bibitem[{\citenamefont{Schubert et~al.}(2005)\citenamefont{Schubert, Weibe,
  Wellin, and Fehske}}]{Schubert}
\bibinfo{author}{\bibfnamefont{G.}~\bibnamefont{Schubert}},
  \bibinfo{author}{\bibfnamefont{A.}~\bibnamefont{Weibe}},
  \bibinfo{author}{\bibfnamefont{G.}~\bibnamefont{Wellin}}, \bibnamefont{and}
  \bibinfo{author}{\bibfnamefont{H.}~\bibnamefont{Fehske}},
  \emph{\bibinfo{title}{HQS@HPC: Comparative numerical study of Anderson
  localisation in disordered electron systems in \em{High Performance computing
  in Science and Engineering, Garching 2004}}} (\bibinfo{publisher}{Springer},
  \bibinfo{year}{2005}).

\end{thebibliography}

\end{document}